\documentclass[twocolumn,aps,pra,showpacs,superscriptaddress,showkeys,twocolumn]{revtex4}
\usepackage{amsmath}
\usepackage{graphicx}

\begin{document}

\title{Vacuum Rabi Splitting in Nanomechanical QED System with Nonlinear
Resonator}

\author{M. Y. Zhao}

\affiliation{College of Applied Sciences, Beijing University of Technology, Beijing,
100124, China}

\author{S. M. Yu}

\affiliation{College of Applied Sciences, Beijing University of Technology, Beijing,
100124, China}

\author{X. Xiao}

\affiliation{College of Applied Sciences, Beijing University of Technology, Beijing,
100124, China}

\affiliation{Department of Physics, Renmin University of China, Beijing, 100872,
China}

\author{Y. B. Gao}
\email[Corresponding author, E-mail: ]{ybgao@bjut.edu.cn}
\affiliation{College of Applied Sciences, Beijing University of Technology, Beijing,
100124, China}

\begin{abstract}
Considering the intrinsic nonlinearity in a nanomechanical resonator
coupled to a charge qubit, vacuum Rabi splitting effect is studied
in a nanomechanical QED (qubit-resonator) system. A driven nonlinear
Jaynes-Cummings model describes the dynamics of this qubit-resonator
system. Using quantum regression theorem and master equation approach,
we have calculated the two-time correlation spectrum analytically.
In the weak driving limit, these analytical results
clarify the influence of the driving strength and nonlinearity parameter
on the correlation spectrum. Also, numerical calculations
confirm these analytical results.
\end{abstract}



\maketitle

\section{introduction}

In quantum optics~\cite{Qoptics} and quantum information,~\cite{QInformation}
the well known Jaynes-Cummings model,~\cite{JC model} one of the most
important models, describes the light-matter interaction between a
two-level quantum system (qubit) and a boson (resonator). Generally
in cavity QED system,~\cite{Cavity QED} nanomechanical QED system~\cite{Nanomechanical QED NJP}
and circuit QED system,~\cite{circuit QED}vacuum Rabi splitting effect
has been used to characterize the coupling strength between a qubit
and a resonator.

Recently, a nanomechanical resonator with frequency of the order of
1 GHz approaches the quantum regime,~\cite{1GHz resonator,Nanomechanical Resonator approaching quantum limit}
it is getting closer to test the basic principles of quantum mechanics.
When a superconducting qubit~\cite{JJ qubit,JJ qubit-Nori} is coupled
to a nanomechanical resonator,~\cite{nanomechanics} we can study quantum
optical properties, such as, quantum decoherence,~\cite{decoherence in Nanomechanical QED,entanglement and decoherence}
Rabi oscillation,~\cite{Rabi oscillation} vacuum Rabi splitting,~\cite{Nanomechanical QED}
classical-quantum transition~\cite{classical-quantum transition-transducer}
and phonon blockade.~\cite{phonon blockade}

Increasing the amplitude of driving, the nonlinearity response of
nanomechanical resonator~\cite{nonlinearity of nanomechanical cantiler}
is not negligible which can be used to detect the classical-quantum
transition.~\cite{classical-quantum transition} When intrinsic nonlinearity
of nanomechanical resonator~\cite{Source of Nonlinearity} is considered
in the qubit-resonator system, we can use superconducting qubit to
probe quantum fluctuations of nonlinear resonator.~\cite{Quantum Heating of a Nonlinear Resonator}
Recently, the nonlinearity can be exploited to generate nonclassical
states in mechanical systems~\cite{nonlinearity to creat nonclassical state,Fock state in mechanical freedom}
and selectively address the nanomechanical qubit transitions in quantum
information processing.~\cite{Qinformation with NR Qubit}

In this paper, the dissipative dynamics of the qubit-resonator system
is solved by master equation approach and quantum regression theorem.~\cite{master equation book}
Comparing with previous results,~\cite{Nanomechanical QED} the influence
of the nonlinearity of nanomechanical resonator on vacuum Rabi splitting
effect is studied analytically and numerically.

The paper is organized as follows. In Sec. II, using a driven nonlinear
Jaynes-Cummings model, we describe the dynamics of a qubit-resonator
system consisting of a superconducting qubit and a nonlinear nanomechanical
resonator. In Sec. III, the two-time correlation spectrum is calculated
analytically. In Sec. IV, vacuum Rabi splitting effect is studied.
Also numerical simulations confirm our analytical results. Finally,
our conclusions are summarized.

\section{Model}

In nanomechanical QED system, consisting a superconducting qubit and
a nanomechanical resonator, we can use a driven Jaynes-Cummings model
to describe the dynamics of this qubit-resonator system,~\cite{Nanomechanical QED}
\begin{eqnarray}
H_{\text{JC}}^{\text{driven}} & = & \omega_{a}\sigma_{+}\sigma_{-}+g\left(a\sigma_{+}+a^{\dagger}\sigma_{-}\right)+\omega_{c}a^{\dagger}a\nonumber \\
 &  & -\xi\sin\left(\omega_{p}t\right)\left(a+a^{\dagger}\right).\label{driven JC Ham}
\end{eqnarray}
Here $H_{\text{JC}}^{\text{driven}}$ is a driven Jaynes-Cummings
type Hamiltonian. The lowering (raising) operator $\sigma_{-}$ ($\sigma_{+}$)
and the annihilation (creation) operator $a$ ($a^{\dagger}$) are
defined to describe the qubit with frequency $\omega_{a}$ and the
resonator with frequency $\omega_{c}$ respectively. The commutation
relations, $[\sigma_{-},\sigma_{+}]=\sigma_{z}$ and $[a,a^{\dagger}]=1$,
are satisfied. The last term in Eq.~(\ref{driven JC Ham}) is a classical
drive with driving constant $\xi$ and driving frequency $\omega_{p}$.
The $g$ denotes the interaction strength between the qubit and the
resonator.

Considering the nonlinearity of nanomechanical resonator, nanomechanical
resonator is not assumed to be an ideal resonator again. Moving into
a frame with rotating frequency $\omega_{p}$, the total Hamiltonian
for this qubit-resonator system writes~\cite{Nanomechanical QED}
\begin{eqnarray}
H_{t} & = & \Delta_{a}\sigma_{+}\sigma_{-}+g\left(a\sigma_{+}+a^{\dagger}\sigma_{-}\right)+\Delta_{c}a^{\dagger}a\nonumber \\
 &  & -\xi\left(a+a^{\dagger}\right)+\chi a^{\dagger}a+\chi\left(a^{\dagger}a\right)^{2},\label{drvien nonlinear JC}
\end{eqnarray}
which describes the dynamics of a driven nonlinear Jaynes-Cummings
model. The linear part $\chi a^{\dagger}a$ and nonlinear part $\chi\left(a^{\dagger}a\right)^{2}$
come from a quartic potential $x^{4}$.~\cite{Source of Nonlinearity}
Here the nonlinearity parameter $\chi$ is small, $\chi\ll g$.

Some parameters in the Hamiltonian Eq.~(\ref{drvien nonlinear JC})
are

\begin{equation}
\Delta_{a}=\omega_{a}-\omega_{p},\ \Delta_{c}=\omega_{c}-\omega_{p}.
\end{equation}
 The detuning $\delta$ between the frequencies of the resonator and
the qubit is
\begin{equation}
\delta=\Delta_{c}-\Delta_{a}.\label{detuning}
\end{equation}

\section{two-time correlation spectrum}

In previous results,~\cite{Nanomechanical QED} the induced electromotive
force between two ends of nanomechanical resonator is
\begin{equation}
V=iBl\sqrt{\frac{\omega_{c}}{2M}}(a^{\dagger}-a).\label{EMF}
\end{equation}
To characterize the vacuum Rabi splitting in this nanomechanical QED
system, the two-time correlation spectrum for the $V$ is
\begin{equation}
S_{V}(\omega)=\frac{1}{\pi}\text{Re}\int_{0}^{\infty}d\tau e^{-i\omega\tau}\left\langle V(\tau)V(0)\right\rangle .\label{correlation spectrum}
\end{equation}
Based on those results in Eqs.~(\ref{EMF},\ref{correlation spectrum}),
the two-time correlation function $\left\langle V(\tau)V(0)\right\rangle $
writes
\begin{eqnarray}
\left\langle V(\tau)V(0)\right\rangle  & \propto & \left\langle a^{\dagger}(\tau)a(0)\right\rangle +\left\langle a(\tau)a^{\dagger}(0)\right\rangle \nonumber \\
 &  & -\left\langle a(\tau)a(0)\right\rangle -\left\langle a^{\dagger}(\tau)a^{\dagger}(0)\right\rangle .\label{two-time correlation}
\end{eqnarray}

Considering the interaction with the environment,
we use the master equation to describe dissipative dynamics of the
qubit-resonator system,~\cite{master equation book} 
\begin{eqnarray}
\dot{\rho} & = & -i\left[H_{t},\rho\right]+\kappa\left(2a\rho a^{\dagger}-a^{\dagger}a\rho-\rho a^{\dagger}a\right)\nonumber \\
 &  & +\frac{\gamma}{2}\left(2\sigma_{-}\rho\sigma_{+}-\sigma_{+}\sigma_{-}\rho-\rho\sigma_{+}\sigma_{-}\right),\label{master equation}
\end{eqnarray}
where the density operator $\rho$
describes the time evolution of the qubit-resonator system. The latter
two terms in Eq.~(\ref{master equation}) describe the decay process
of the resonator and the qubit respectively. The parameters $\kappa$
and $\gamma$ denote the decay rates of the resonator
and the qubit. Here the Markov approximation is satisfied.

In this case, the number operator
$N=\sigma_{+}\sigma_{-}+a^{\dagger}a$ is defined
to characterize the total number of excitations of the qubit-resonator
system. The operator $N$ satisfies 
\[
N\left\vert j,k\right\rangle =\left(j+k\right)\left\vert j,k\right\rangle ,
\]
where $\left\vert j,k\right\rangle =\vert j\rangle\otimes\vert k\rangle$
is the state of the qubit-resonator system, the index $j$ ($k$)
denotes the state of the qubit (resonator) and satisfies $\sigma_{+}\sigma_{-}|j\rangle=j|j\rangle$
($a^{\dagger}a|k\rangle=k|k\rangle$) for $j=0,1$ ($k=0,1,...$).

In this paper, the weak driving limit is adopted,
$\xi\rightarrow0.$ When time approaches infinite, $t\rightarrow\infty$,
the qubit-resonator system will decay into the vacuum state $\vert00\rangle$.
The weak driving $\xi$ will induce the transitions from the vacuum
state $\vert00\rangle$ to the other excited states ($\vert01\rangle$
and $\vert10\rangle$). Under the lowest order perturbation theory,
we need to consider only one-phonon excitation in the qubit-resonator
system,
\begin{equation}
N=j+k=0,\,\,1.
\end{equation}
The Hilbert space for the reduced density matrix $\rho$ in Eq.~(\ref{master equation})
reduces into a smaller subspace with a truncated basis
\begin{equation}
\left\{ \left\vert j,k\right\rangle ,\,\,j+k=0,\,\,1\right\} .\label{truncated basis}
\end{equation}

Thus, in this truncated basis, the corresponding density matrix elements
satisfying the master equation in Eq.~(\ref{master equation}) are

\begin{eqnarray}
\frac{d\rho_{00,00}}{d\tau} & = & i\xi\rho_{01,00}-i\xi\rho_{00,01}\nonumber \\
 &  & +2\kappa\rho_{01,01}+\gamma\rho_{10,10},\nonumber \\
\frac{d\rho_{00,01}}{d\tau} & = & \left[i\left(\Delta_{c}+2\chi\right)-\kappa\right]\rho_{00,01}\nonumber \\
 &  & +ig\rho_{00,10}+i\xi\left(\rho_{01,01}-\rho_{00,00}\right),\nonumber \\
\frac{d\rho_{00,10}}{d\tau} & = & \left(i\Delta_{a}-\frac{\gamma}{2}\right)\rho_{00,10}+ig\rho_{00,01}+i\xi\rho_{01,10},\nonumber \\
\frac{d\rho_{01,01}}{d\tau} & = & -2\kappa\rho_{01,01}+ig\left(\rho_{01,10}-\rho_{10,01}\right)\nonumber \\
 &  & +i\xi\left(\rho_{00,01}-\rho_{01,00}\right),\nonumber \\
\frac{d\rho_{01,10}}{d\tau} & = & \left[-i\left(\delta+2\chi\right)-\kappa-\frac{\gamma}{2}\right]\rho_{01,10}\nonumber \\
 &  & +ig\left(\rho_{01,01}-\rho_{10,10}\right)+i\xi\rho_{00,10},\nonumber \\
\frac{d\rho_{10,10}}{d\tau} & = & -\gamma\rho_{10,10}+ig\left(\rho_{10,01}-\rho_{01,10}\right).\label{elements}
\end{eqnarray}
In the long-time limit, the system stays in a steady state, we can
take $\rho_{00,00}\sim1$ and $\rho_{00,00}\gg\rho_{01,01}\left(\rho_{10,10}\right)$.
Here can see that $\rho_{00,01}$ ($\rho_{00,10}$) scales as the
order of $\xi$ and $\rho_{01,01}\left(\rho_{10,10}\right)$ scales
as the order of $\xi^{2}$. Keeping terms of the $\xi$ and dropping
the higher terms of the $\xi^{2}$, we get
\begin{eqnarray}
\frac{d\rho_{00,01}}{d\tau} & = & \left[i\left(\Delta_{c}+2\chi\right)-\kappa\right]\rho_{00,01}+ig\rho_{00,10}-i\xi\rho_{00,00},\nonumber \\
\frac{d\rho_{00,10}}{d\tau} & = & \left(i\Delta_{a}-\frac{\gamma}{2}\right)\rho_{00,10}+ig\rho_{00,01}.\label{first order perturbation}
\end{eqnarray}

Applying Laplace transformation, we have solved
\begin{equation}
\rho_{00,01}\left(\tau\right)=\mu\left(\tau\right)\rho_{00,01}^{ss}+\nu\left(\tau\right)\rho_{00,10}^{ss}-i\xi C\left(\tau\right)\rho_{00,00}^{ss}.\label{solution for 1st-order}
\end{equation}
Here the coefficients ($\mu\left(\tau\right)$, $\nu\left(\tau\right)$
and $C\left(\tau\right)$) are dependent of time, i.e.,
\begin{eqnarray*}
\mu\left(\tau\right) & = & \frac{\Gamma_{1}+i\omega_{1}+\left(i\Delta_{a}-\frac{\gamma}{2}\right)}{2iG}e^{-\Gamma_{1}\tau}e^{-i\omega_{1}\tau}\\
 &  & +\frac{\Gamma_{2}+i\omega_{2}+\left(i\Delta_{a}-\frac{\gamma}{2}\right)}{-2iG}e^{-\Gamma_{2}\tau}e^{-i\omega_{2}\tau},\\
\nu\left(\tau\right) & = & \frac{g}{-2G}e^{-\Gamma_{1}\tau}e^{-i\omega_{1}\tau}+\frac{g}{2G}e^{-\Gamma_{2}\tau}e^{-i\omega_{2}\tau},\\
C\left(\tau\right) & = & \frac{1}{2iG}\frac{\Gamma_{1}+i\omega_{1}+\left(i\Delta_{a}-\frac{\gamma}{2}\right)}{\Gamma_{1}+i\omega_{1}}\left(1-e^{-\Gamma_{1}\tau}e^{-i\omega_{1}\tau}\right)\\
 &  & -\frac{1}{2iG}\frac{\Gamma_{2}+i\omega_{2}+\left(i\Delta_{a}-\frac{\gamma}{2}\right)}{\Gamma_{2}+i\omega_{2}}\left(1-e^{-\Gamma_{2}\tau}e^{-i\omega_{2}\tau}\right),
\end{eqnarray*}
\begin{equation}
\Gamma_{n}=\frac{1}{2}\left(\frac{\gamma}{2}+\kappa\right)+\left(-1\right)^{n}\text{Im}G,\label{decay rate}
\end{equation}
\begin{equation}
\omega_{n}=\left(-1\right)^{n+1}\text{Re}G-\frac{1}{2}\Delta_{a}-\frac{1}{2}\left(\Delta_{c}+2\chi\right),\label{peak frequency}
\end{equation}
and
\[
G=\sqrt{g^{2}-\frac{1}{4}\left[i\left(\delta+2\chi\right)+\left(\frac{\gamma}{2}-\kappa\right)\right]^{2}}
\]
for $n=1,2$.

From the above results in Eq.~(\ref{first order perturbation}), we
can calculate the single-time function
\begin{equation}
\left\langle a(\tau)\right\rangle =Tr\left(a\left(0\right)\rho\left(\tau\right)\right)=\rho_{01,00}\left(\tau\right)\label{single-time}
\end{equation}
where $\rho\left(0\right)=\rho^{ss}$ and the index $ss$ means the
steady state. Through some simple calculations, we obtain the steady
solution of the density matrix element $\rho_{01,01}\left(\rho_{10,10}\right)$,
\begin{eqnarray}
\rho_{00,01}^{ss} & = & \frac{\xi}{\left(\Delta_{c}+2\chi\right)+i\kappa-\frac{g^{2}}{\left(\Delta_{a}+i\frac{\gamma}{2}\right)}},\nonumber \\
\rho_{00,10}^{ss} & = & -\frac{g}{\left(\Delta_{a}+i\frac{\gamma}{2}\right)}\rho_{00,01}^{ss}.\label{steady state}
\end{eqnarray}

Using the quantum regression theorem~\cite{Nanomechanical QED,master equation book,Xiao99},
we can obtain the two-time correlation function
\begin{equation}
\left\langle a(\tau)a^{\dagger}(0)\right\rangle =\text{Tr}\left\{ a(\tau)a^{\dagger}(0)\rho\left(0\right)\right\} =\mu\left(\tau\right)^{*}.\label{2nd order term-1}
\end{equation}
The other two-time functions can be obtained as
\begin{equation}
\left\langle a^{\dagger}(\tau)a(0)\right\rangle =-i\xi C\left(\tau\right)\rho_{01,00}^{ss},\label{2nd order term-2}
\end{equation}
\begin{equation}
\left\langle a^{\dagger}(\tau)a^{\dagger}(0)\right\rangle =0\label{2nd order term-3}
\end{equation}
and
\begin{equation}
\left\langle a(\tau)a(0)\right\rangle =0.\label{2nd order term-4}
\end{equation}
Based on the above results in Eqs.~(\ref{2nd order term-1},\ref{2nd order term-2},\ref{2nd order term-3},\ref{2nd order term-4}),
the two-time correlation function for the induced electromotive force
is obtained,
\begin{equation}
\left\langle V(\tau)V(0)\right\rangle \propto\mu\left(\tau\right)^{*}.\label{correlation}
\end{equation}
In the limit of weak driving, we can neglect the terms ($i\xi\rho_{01,00}\left(0\right)$,
$\rho_{02,00}\left(0\right)$, $\rho_{11,00}\left(0\right)$, $\rho_{01,01}\left(0\right)$
and $\rho_{01,10}\left(0\right)$) which are proportional to the $\xi^{2}$
in Eq. (\ref{correlation}).

Using the formula,
\begin{eqnarray*}
\int_{0}^{\infty}d\tau e^{-i\left(\omega-\omega_{0}\right)\tau-\Gamma\tau} & = & \frac{\Gamma-i\left(\omega-\omega_{0}\right)}{\left(\omega-\omega_{0}\right)^{2}+\Gamma^{2}},
\end{eqnarray*}
the correlation spectrum in Eq.~(\ref{correlation spectrum}) is calculated
as
\begin{equation}
S_{V}(\omega)\propto\frac{\text{Re}\eta_{1}}{\left(\omega-\omega_{1}\right)^{2}+\Gamma_{1}^{2}}+\frac{\text{Re}\eta_{2}}{\left(\omega-\omega_{2}\right)^{2}+\Gamma_{2}^{2}}.\label{spectrum-analytic}
\end{equation}
Where some parameters are defined,
\[
\eta_{n}=\frac{\Gamma_{n}-i\omega_{n}+\left(-i\Delta_{a}-\frac{\gamma}{2}\right)}{\left(-1\right)^{n}2iG^{*}}\left[\Gamma_{n}-i\left(\omega-\omega_{n}\right)\right]
\]
for $n=1,\ 2$.

\section{vacuum rabi splitting}

Generally in vacuum Rabi splitting, the splitting frequency $\Delta\omega$
provides the information of the coupling $g$ between the qubit and
the resonator. As seen in Eq.~(\ref{spectrum-analytic}), nonlinearity
parameter $\chi$ modifies the decay rate $\Gamma_{1}$ ($\Gamma_{2}$)
and central frequency $\omega_{1}$ ($\omega_{2}$) of two peaks in
the spectrum $S_{V}\left(\omega\right)$. And in the limit of weak
driving, the $\xi$ does not affect the spectrum $S_{V}\left(\omega\right)$,
it means that we can use this driven nonlinear Jaynes-Cummings model
to characterize vacuum Rabi splitting effect very well. The corresponding
splitting frequency between two peaks in the spectrum $S_{V}(\omega)$
is
\begin{equation}
\Delta\omega=\vert\text{\ensuremath{\omega}}_{1}-\omega_{2}\vert=2\text{Re}G\label{splitting}
\end{equation}
which is independent of the driving strength $\xi$ and determined
by the couple strength $g$, the detuning $\delta$, decay rate $\kappa$
($\gamma$) and nonlinearity parameter $\chi$.

Assuming nanomechanical resonator as an ideal resonator, $\chi=0$,
the spectrum $S_{V}\left(\omega\right)$ in Eq.~(\ref{spectrum-analytic})
will be same as the previous results.~\cite{Nanomechanical QED} When
the resonant condition ($\delta=0$) and the strong-coupling limit
($g>>\left[\kappa,\gamma\right]$) are adopted, we obtain the well-known
splitting frequency~\cite{Cavity QED}
\begin{equation}
\Delta\omega\simeq2g.\label{splitting-JC}
\end{equation}

To further clarify the dependence of correlation spectrum $S_{V}(\omega)$
on nonlinearity parameter $\chi$ and the driving strength $\xi$
more clearly, the resonant condition is adopted, $\Delta_{a}=\Delta_{c}=1.0$.
The other parameters are $g=0.2,\ \kappa=0.004,\ \gamma=0.004$ and
$1$ GHz is taken as the unit for all these parameters.~\cite{Nanomechanical QED,Qinformation with NR Qubit,parameters}

\begin{figure}[ht]
\includegraphics[width=8cm]{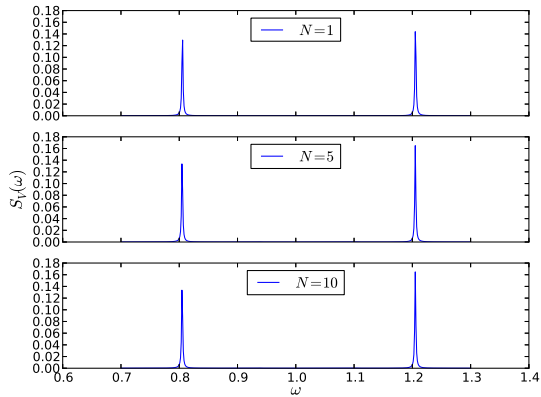}

\caption{Correlation spectrum $S_{V}\left(\omega\right)$ versus $\omega$
for different values of excitations, we take $N=1,\ 5,\ 10$. The other
parameters are $\chi=0.01\ ,\xi=0.02$.}

\label{different N}
\end{figure}

Numerical calculations by QuTiP~\cite{python} are illustrated with
some plots in the following. In Fig.~\ref{different N}, we choose the
values of $\xi=0.02$ and $\chi=0.01$, the increasing of the number
of total excitations $N$ tells that the value of $N=10$ is suitable
in the other plots of the spectrum $S_{V}(\omega)$.

\begin{figure}[th]
\includegraphics[width=8cm]{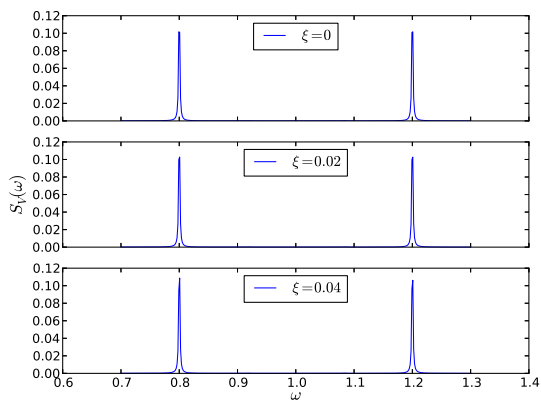}\caption{Correlation spectrum $S_{V}\left(\omega\right)$ versus $\omega$
for different values of the driving strength, we take $\xi=0,\ 0.02,\ 0.04$.
The other parameters are $\chi=0$.}
\label{driving strength}
\end{figure}

Figure~\ref{driving strength} shows, in the limit of weak driving
($\xi\ll g$), there are two peaks in the spectrum and the increasing
of the driving strength does not affect the spectrum obviously, which
confirms the analytical calculations in Eq.~(\ref{spectrum-analytic}).
Then we take the value of $\xi=0.02$ in the following plots.

\begin{figure}[ht]
\includegraphics[width=8cm]{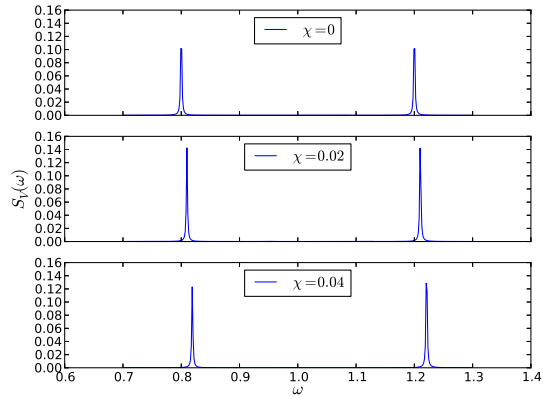}
\caption{Correlation spectrum $S_{V}\left(\omega\right)$ versus $\omega$
for different values of the driving strength, we take $\chi=0,\ 0.02,\ 0.04$.
The other parameters are $\xi=0$.}
\label{nonlinearity parameter}
\end{figure}

As shown in Fig.~\ref{nonlinearity parameter}, the increasing of
nonlinearity parameter leads to the shift of central frequency of
two peaks dramatically and does not change the distance between the
centers of two peaks obviously because nonlinearity parameter is much
smaller than the coupling $g$, we can see it in Eq.~(\ref{splitting}).

Figure~\ref{driving strength} and Figure~\ref{nonlinearity parameter}
demonstrate the dependence of the spectrum $S_{V}\left(\omega\right)$
on the driving strength $\xi$ and nonlinearity parameter $\chi$.
These numerical results agree with those analytical results in Eq.~(\ref{spectrum-analytic}),
both of them maintain that the weak driving strength $\xi$ does not
change the splitting frequency $\Delta\omega$, nonlinearity parameter
$\chi$ changes the heights of two peaks ($\text{Re}\eta_{1}$ and
$\text{Re}\eta_{2}$) and the shifts of central frequency ($\omega_{1}$
and $\omega_{2}$) obviously.

\section{conclusions}

In this paper, vacuum Rabi splitting effect is studied to provided
the information of the coupling $g$. Considering the intrinsic nonlinearity
in nanomechanical resonator, a driven nonlinear Jaynes-Cummings model
is used to describe the dynamics of the qubit-resonator system. Using
quantum regression theorem, the dissipative dynamics of the qubit-resonator
system is solved by master equation approach. Here, the two-time correlation
spectrum is analytically calculated to clarify the dependence of correlation
spectrum on the driving strength $\xi$ and nonlinearity parameter
$\chi$. Because of small nonlinearity in nanomechanical resonator,
we find that nonlinearity parameter leads to the shifts of central
frequency ($\omega_{1}$ and $\omega_{2}$) and does not change the
splitting frequency $\Delta\omega$ obviously. In Fig.~\ref{driving strength}
and Fig.~\ref{nonlinearity parameter}, numerical results plotted by
QuTiP agree with the analytical results in Eq.~(\ref{spectrum-analytic}).
\begin{acknowledgments}
We thank the discussions with Professor P. Zhang.\end{acknowledgments}

\end{document}